\documentstyle[10pt,conf-X]{article}
\input psfig
\begin{document} 
\small
\heading{
\centerline{The Cosmic Foam and the Self-Similar Cluster Distribution}\\
}
\author{Rien van de Weygaert}
\vspace{-0.25truecm}
\address{Kapteyn Institute, University of Groningen, Groningen, the Netherlands}
\vspace{-0.75truecm}
\section{Foamlike Patterns in Cosmic Structure}
\vspace{-0.25truecm}
Probing cosmic large scale structure on the basis of X-ray observations 
puts particular emphasis on the densest regions within the global 
matter distribution, the rich clusters of galaxies. Understanding
the relationship between the cluster distribution and the underlying 
matter distribution is therefore a key element in any assessment of 
cosmic structure on the basis of samples of X-ray selected clusters. 
By now, the foamlike arrangement of matter and galaxies on Megaparsec 
scales has become a well-established feature of the cosmic matter 
distribution, consisting of an assembly of anisotropic elements, filaments 
and walls of various sizes, surrounding large underdense void regions 
and sprinkled with wholly or partially virialized dense clumps of 
matter, varying in size from rich clusters of galaxies down to small 
groups of a few galaxies. It is one of successes of gravitational instability 
cosmic structure formation theories to find that this pattern of 
walls, filaments and voids appears to be the generic outcome of 
these scenarios. 

A major obstacle in quantifying this quintessential aspect of cosmic 
structure is the lack of a systematic insight into the dynamical and 
statistical aspects of cellular geometries as well as the absence of 
a readily available and well-established mathematical machinery to 
evaluate and compare observations and simulations. Stochastic geometry 
-- the branch of mathematics concerned with nontrivial geometrical 
concepts involving stochastic behaviour of one or more of their 
characteristics -- may be expected to contribute significantly to 
further such understanding. 

\begin{figure}[t]
\vspace{-2.0truecm}
\hbox{\hskip -1.0truecm\psfig{figure=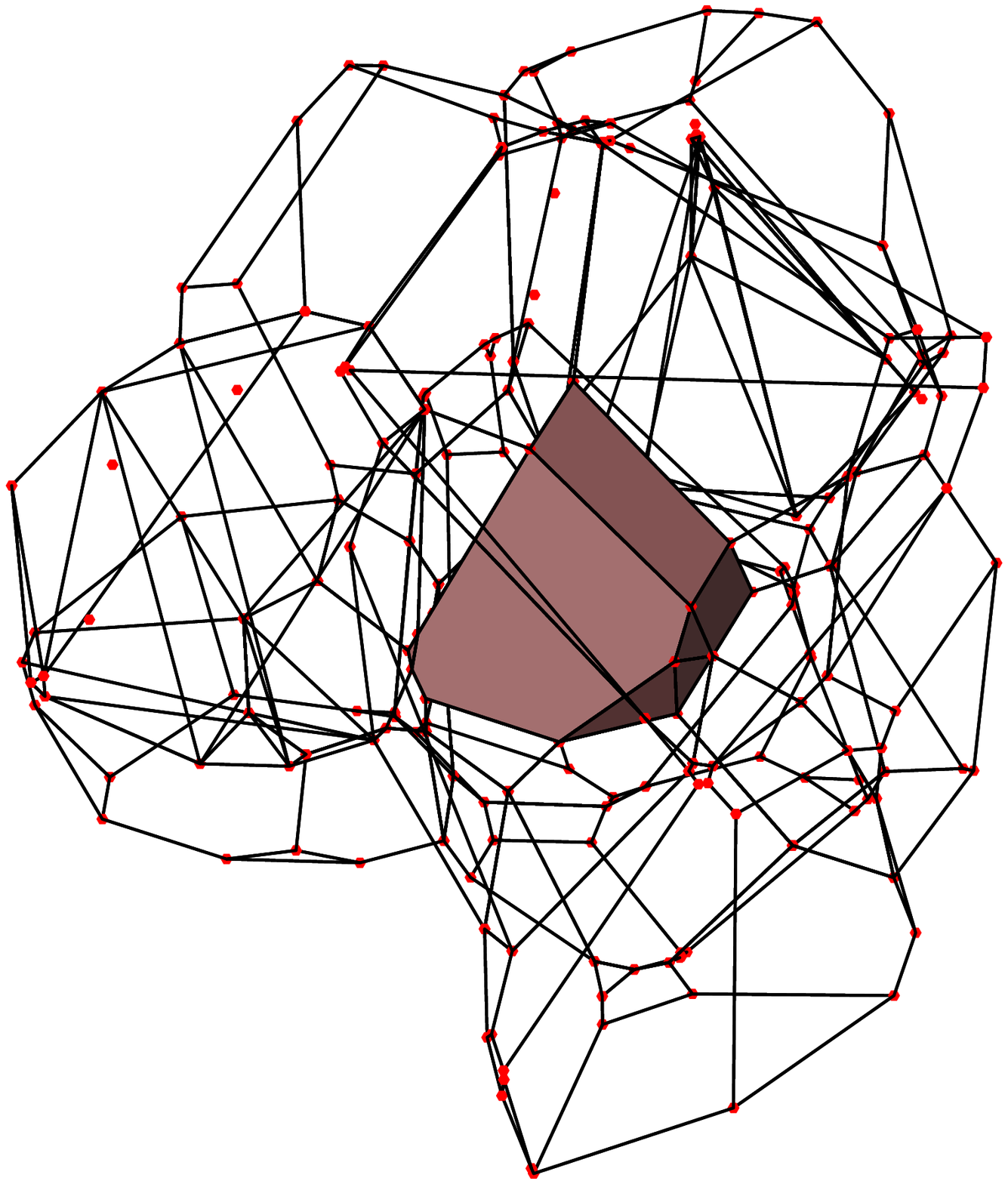,height=7.0cm}}
\vskip -7.0truecm
\hbox{\hskip 6.truecm\psfig{figure=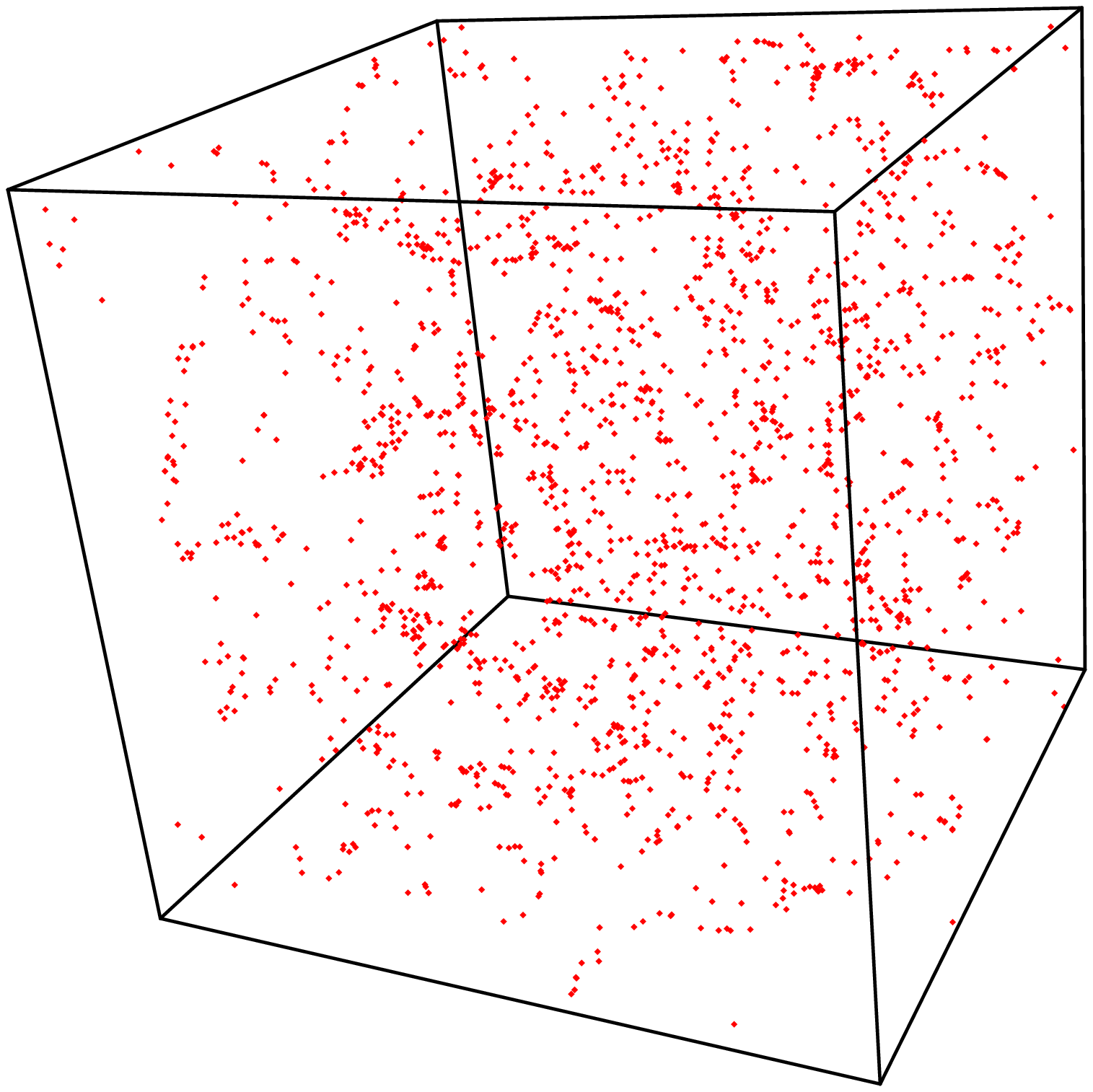,height=7.0cm}}
\caption[]{A clump of Voronoi polyhedra from a tessellation of 1000 cells. 
One cell is surface-shaded, the surrounding neighbouring 
cells are depicted in wire-frame fashion. The corresponding vertices 
are indicated by solid dots. Distribution of all Voronoi vertices 
within a cubic volume. Courtesy: Jacco Dankers.}
\vspace{-0.25truecm}
\end{figure}
\vspace{-0.50truecm}
\section{Voronoi Tessellations}
\vspace{-0.25truecm}
The canonical example of a stochastic geometrical model for a cellular 
division of space is that of Voronoi tessellations. This space filling 
network of convex polyhedra offers a surprisingly realistic and versatile 
representation of the characteristics and features of the foamlike or cellular 
spatial arrangement of matter in the Universe. In short, it is defined 
through a spatial  distribution of nuclei\cite{W91}. Each nucleus 
corresponds to one Voronoi cell, which 
comprises that part of space closer to this nucleus than to any of the 
other. The walls and edges forming the polyhedrons' surfaces are 
identified with the wall-like and filamentary superclusters in the 
galaxy distribution, the vertices with and massive clusters of galaxies, 
while the interior of the Voronoi cells corresponds to 
the large void regions barren of galaxies. The morphology of artificial 
galaxy distributions set up within this network bears a striking resemblance 
to that observed in the more complicated circumstances of the real Cosmos 
or that of the artificial reality of computer simulations of structure 
formation. Figure 1 shows a conglomerate of several neighbouring Voronoi 
cells, with one of the 
cells by shaded surface and a wire-framed representation in the case of 
its neighbouring cells. The solid lines are the edges (filaments) in 
the network, the dots (coloured red) are the vertices of the Voronoi 
tessellation. In principle a Voronoi tessellation could be regarded as 
a mere geometric toy model, a heuristic geometrical description of 
a substantially more complicated reality. However, a detailed 
assessment of the role of voids in structure formation provides  
a physical ground for them representing reality, whether the observed 
or simulated one, in a more subtle way\cite{I84}\cite{D93}. This is 
certainly reinforced by their striking resemblance to the observed or 
simulated foamlike appearance of the galaxy and matter distribution.

\vspace{-0.50truecm}
\section{Voronoi Vertex Clustering}
\vspace{-0.25truecm}
Even in the limited setting of Figure 1 it is evident that the vertex 
distribution is not a random Poisson distribution. The full spatial 
distribution of Voronoi vertices in the full cubic volume (righthans frame 
fig. 1) clearly 
involves a substantial degree of clustering. This impression of strong 
clustering, on scales smaller than or of the order of the cellsize 
$\lambda_{\rm C}$, is most evidently confirmed by their two-point correlation 
function $\xi(r)$. Not only can we discern a clear positive signal but -- 
surprising at the time of its finding on the basis of similar computer 
experiments\cite{WI89} -- out to a distance 
of at least $r \approx 1/4\,\lambda_{\rm C}$ the correlation function 
appears to be an almost perfect power-law, 
\begin{equation}
\xi(r)\,=\,(r_{\rm o}/r)^{\gamma}\,, 
\end{equation}
with a slope $\gamma \approx 1.9-2.0$. Its amplitude, traditionally 
expressed in terms of the ``clustering length'' $r_{\rm o}$, at which
$\xi(r_{\rm o})\,=\,1\,$, has a value 
$r_{\rm o} \approx 0.29\,\lambda_{\rm C}$. Beyond this range, the power-law 
behaviour breaks down and following a 
gradual decline the correlation function rapidly falls of to a zero value once 
distances are of the order of the cellsize. However, rather than a 
characteristic geometric scale, $r_{\rm o}$ is more a measure for 
the ``compactness'' of the spatial clustering, set mainly by the 
small-scale clustering. A more significant scale within the context of 
the geometry of the spatial patterns in the density distribution is the 
``correlation length'' $r_{\rm a}$, the scale at which 
$\xi(r_{\rm a})\,=\,0\,$. As a genuine scale of coherence, it is 
more relevant to the morphology of the nontrivial spatial structures 
we seek to study. Beyond $r_{\rm a}$ the distribution of Voronoi vertices 
is practically uniform. 

If we interpret the clustering length $r_{\rm o} \approx 20h^{-1}\,
\hbox{Mpc}$, usually found for samples of rich clusters of 
galaxies\cite{BS83}, within the context of a Voronoi tessellation it 
would imply a cellsize of $\lambda_{\rm C} \approx 70h^{-1}\,\hbox{Mpc}$. 
Although the two-point cluster-cluster correlation function reproduced 
by the Voronoi vertices fits very well to the function yielded from the 
observations, the large cell size $\lambda_c$ may be a complication. 
It is surely well in excess of the $25h^{-1}-35h^{-1}\,\hbox{Mpc}$ size of the 
voids in the galaxy distribution. Moreover, also 
within the Voronoi concept itself it would conflict with the 
clustering of objects dwelling in the walls and filaments of the 
same tessellation framework. Clustering analysis of 
such configurations\cite{W91} reveals that the two-point 
correlation function of galaxies confined to the walls -- as well as 
for those confined to the edges -- also displays distinct power-law 
behaviour at sub-cellular scales. The  involved clustering length, 
however, is different from that of the vertices in the same framework. For 
the wall galaxies it is but half the value of that of the vertices, 
$r_{w,{\rm o}} \approx 0.14\,\lambda_{\rm C}$\cite{W91},\cite{W99}. 
If $r_{\rm w,o}$ is identified with the galaxy-galaxy clustering length 
of $r_{\rm g,o} \approx 5h^{-1}\,\hbox{Mpc}$, this would yield a 
cellsize of $\approx 35h^{-1}\,\hbox{Mpc}$. The latter is suggestively 
similar to the size of the actually observed voids, which may be a tantalizing 
hint for a profound relationship between the clustering length 
$r_{\rm o}$ and the typical cellular scale of the cosmic foam network.

Adressing this apparent inconsistency between vertex and wall clustering 
we first observe that the vertex correlation function of eqn.~(1) concerns 
the whole sample of vertices, irrespective of any possible selection 
effects based on one or more relevant physical aspects. In reality, it will 
be almost inevitable to invoke some sort of biasing through the defining 
criteria of the catalogue of clusters. Interpreting the Voronoi model in 
its quality of asymptotic approximation to the galaxy distribution, its 
vertices will automatically comprise a range of ``masses''. Neglecting 
the details of the temporal  
evolution, we may assign each Voronoi vertex a ``mass'' equal to the 
total amount 
of matter that ultimately will flow towards that vertex. Applying the 
``Voronoi streaming model''\cite{W91} as a reasonable description of the 
clustering process, it is reasonably straightforward if cumbersome to 
calculate the ``mass'' or ``richness'' ${\cal M}_{\rm V}$ of each 
Voronoi vertex by pure geometric means\cite{W99}. It concerns the volume 
of a non-convex polyhedron centered on the Voronoi vertex, with 
related Voronoi nuclei as one of the polyhedral vertices. These nuclei 
are the ones that supply the Voronoi vertex with inflowing matter. 
Evidently, the ``mass'' is larger for vertices on the 
surface of large Voronoi cells.
 
\begin{figure}[t]
\vspace{-1.truecm}
\hbox{\hskip -2.0truecm\psfig{figure=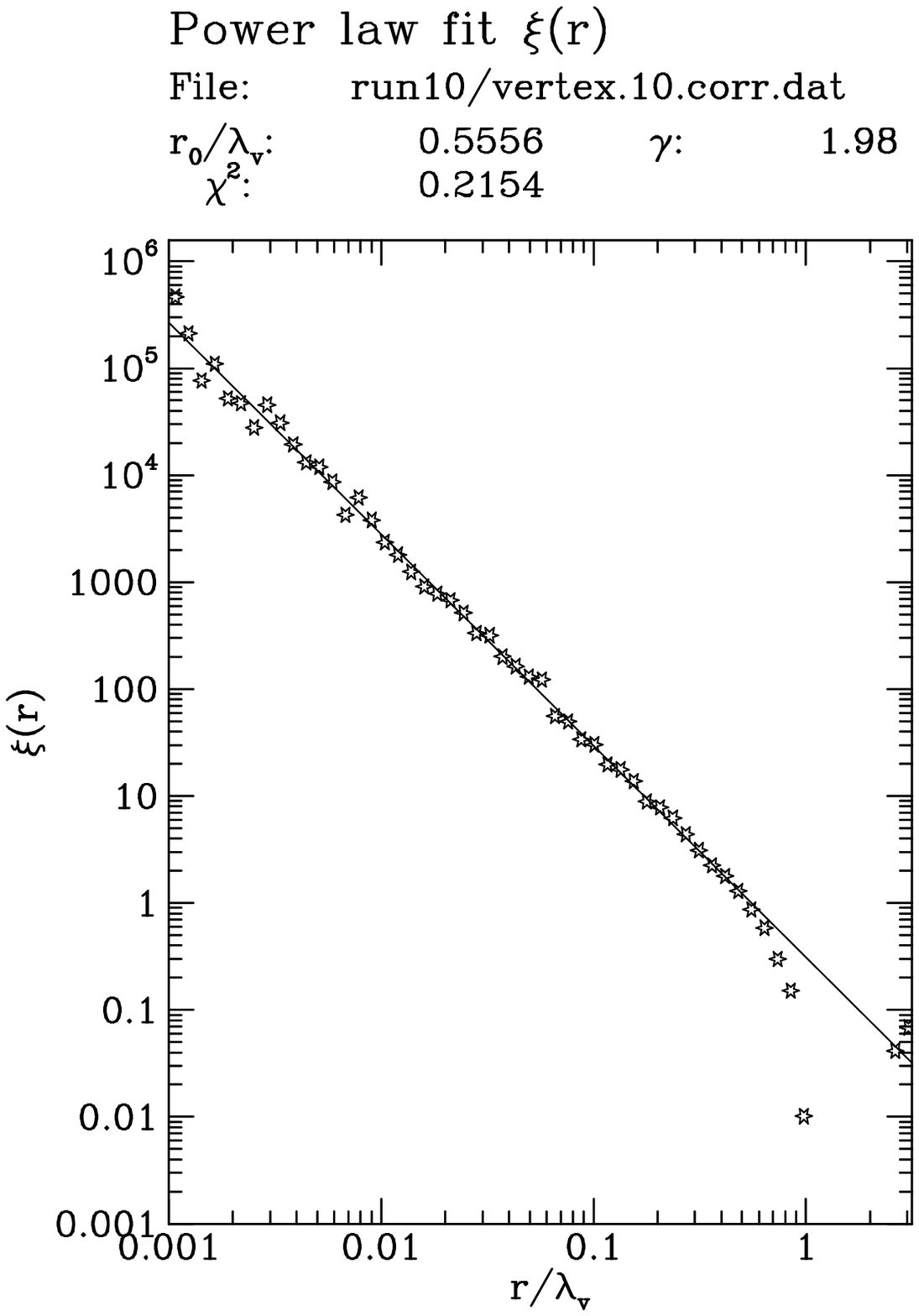,height=4.0cm}}
\vskip -4.5truecm
\hbox{\hskip 0.5truecm\psfig{figure=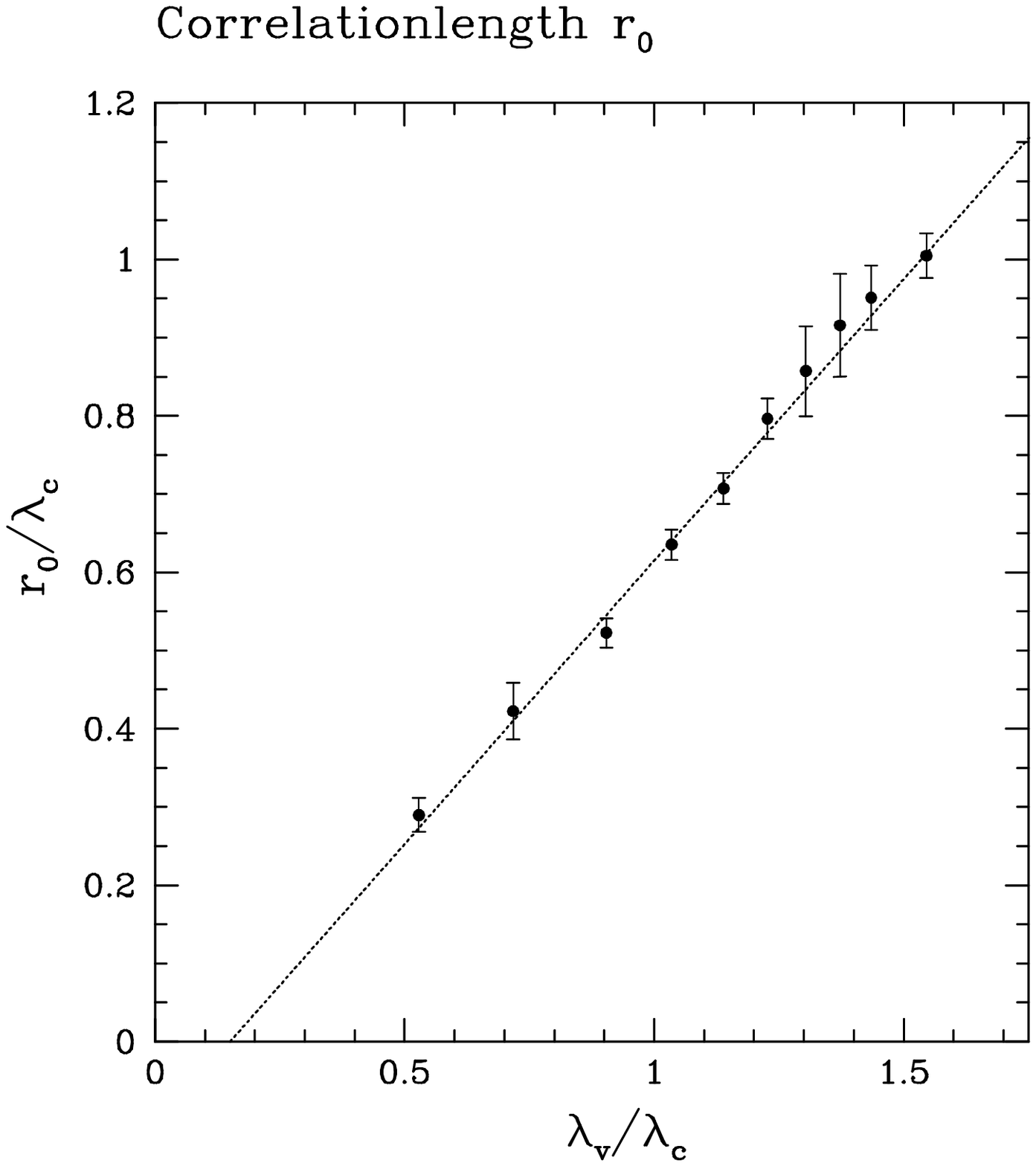,height=7.5cm}}
\vskip -7.5truecm
\hbox{\hskip 6.0truecm\psfig{figure=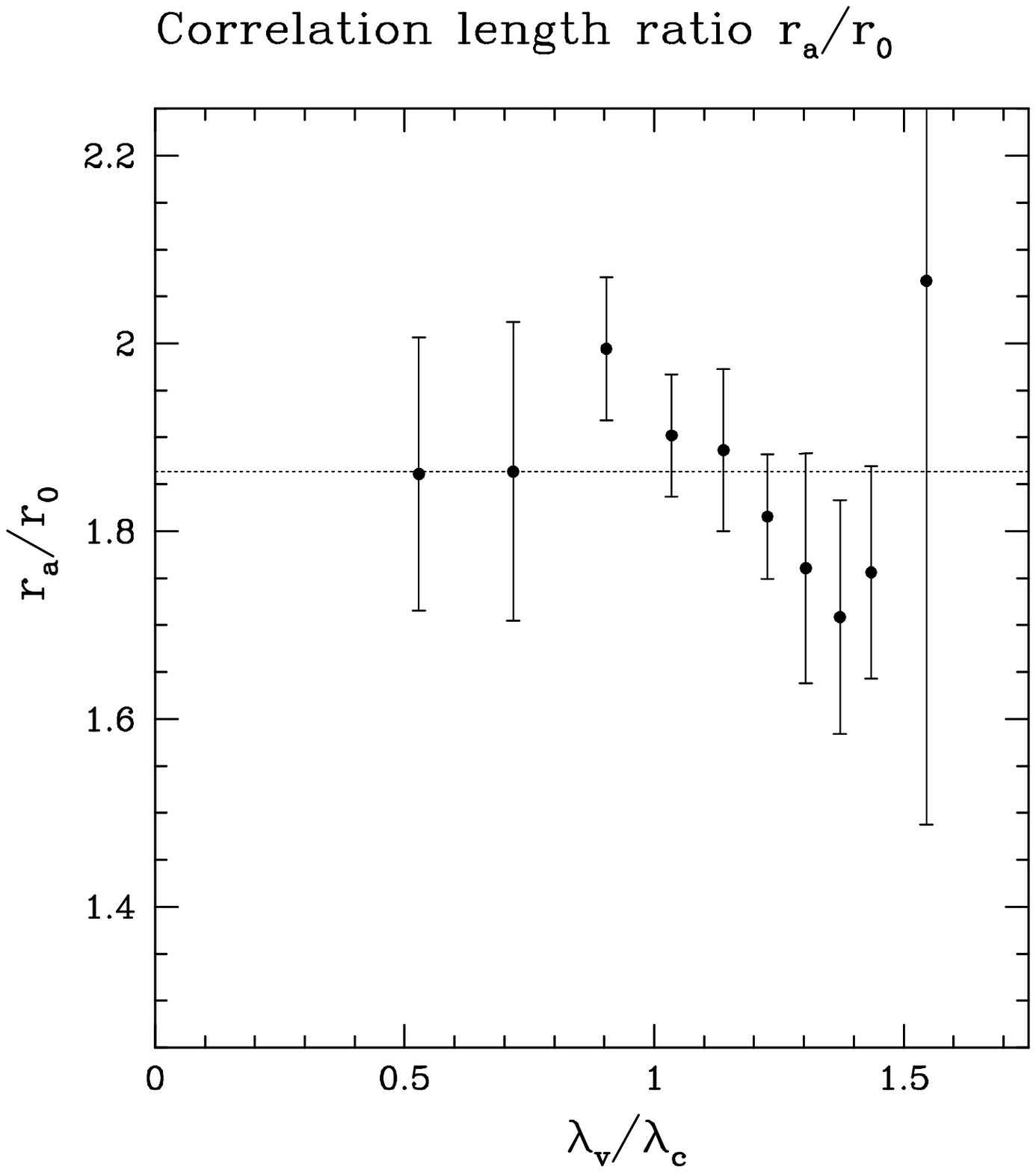,height=7.5cm}}
\vskip -1.0truecm
\caption[]{Voronoi tessellation scaling. Left insert: Two-point correlation 
function $\xi(r)$ Voronoi vertices. The clustering length $r_{\rm o}$ 
(left) and the ratio $r_{\rm a}/r_{\rm o}$ (right) as a function of average 
vertex distance.}
\vspace{-0.25truecm}
\end{figure}
The vertex samples in our study consist of all Voronoi 
vertices with a richness equal to higher than a limiting value,  
the ``sample richness'' ${\cal M}_{\rm S}$\cite{W99}. Subsequently, we 
determined the correlation function characteristics for each of the 
vertex samples. Assessing their behaviour as a function of the 
average distance $\lambda_{\rm V}$ between the sample vertices, 
for samples encompassing 
from $10\%$ up to $100\%$ of the total number of Voronoi vertices and 
thus for $\lambda_{\rm V} \approx 0.5-1.5 \times \lambda_{\rm C}$, 
revealed a remarkable and tantalizing scaling of the clustering 
phenomenon\cite{W99}. 

All subsamples of Voronoi vertices have a two-point correlation function 
that out to a certain range retains a power-law behaviour almost 
exactly similar the one of the full sample of vertcies. No significant 
difference in power law slope between the various subsamples can be 
discerned, all have $\gamma \sim 1.8-2.0$. On the 
other hand, both clustering length $r_{\rm o}$ and correlation length 
$r_{\rm a}$ do display a systematic dependence on sample richness. 
Both $r_{\rm o}$ and $r_{\rm a}$ increase proportionally to the 
sample richness, and hence the average vertex distance $\lambda_{\rm V}$ in 
the samples ! The more massive the vertices are, the more strongly they 
cluster. In other words, in the line of the random field ``peak 
biasing'' scheme\cite{K84}, we here find a purely geometrically based 
biasing scheme. Interesting in this respect is to observe that the increase of 
$r_{\rm V,o}$ is perfectly {\it linearly} proportional to the mutual vertex 
distance $\lambda_{\rm V}$ in the sample (Fig. 2, left frame). 
This suggests that Voronoi vertex clustering is a perfect 
realization of the clustering scaling description proposed by 
Szalay \& Schramm\cite{SS85}. It also adheres to the increasing level of 
clustering that selections of more massive clusters appear to display 
in large-scope N-body simulations\cite{C98} although there are  
telling differences in detailed behaviour. 

For our purpose, even more significant is the uncovered scaling of 
the correlation length $r_{\rm V,a}$, similar in character to that 
of $r_{\rm V,o}$. The increase of $r_{\rm a}$ also turns out to be 
almost exactly linearly proportional to the average vertex distance. 
The repercussions are manyfold. It means that the selected vertices still 
have a strong positive correlation at scales where the poorer samples do 
not possess any clustering. Moreover, samples with more massive 
clusters are expected to have a clustering that extends out further than 
that of their poorer brethren. This may be a significant observation in 
the light of the finding that samples of rich galaxy clusters seem to 
have a positive correlation at scales of tens of Megaparsec, quite in 
excess of scales with appreciable galaxy clustering. Finally, the 
implied constant ratio between clustering and correlation length, 
$r_{\rm a}/r_{\rm o} \approx 1.86$ (Fig. 2, right frame), implies 
a perfect self-similar scaling of the Voronoi vertex distribution. The 
complete correlation function $\xi_s(r)$ of each selected subsample $s$, 
and not just the part in the power-law range, is a self-similar mapping of 
an elementary function $\xi_{\rm el}$ scaled by means of a 
characteristic lengthscale parameter $L_s$. 

\vspace{-0.50truecm}
\section{Bias and Cosmic Geometry: Conclusions}
\vspace{-0.25truecm}
The above results form a tantalizing indication 
for the existence of self-similar clustering behaviour in spatial 
patterns with a cellular or foamlike morphology. It might hint at an 
intriguing and intimate relationship between the cosmic foamlike geometry 
and a variety of aspects of the spatial distribution of galaxies and 
clusters. One important implication is that with clusters residing at a
subset of nodes in the cosmic cellular framework, a configuration 
certainly reminiscent of the observed reality, it would explain 
why the level of clustering of clusters of galaxies becomes stronger 
as it concerns samples of more massive clusters. In addition, it would 
succesfully reproduce positive clustering of clusters over scales 
substantially exceeding the characteristic scale of voids and other 
elements of the cosmic foam. At these Megaparsec scales there is 
a close kinship between the measured galaxy-galaxy two-point 
correlation function and the foamlike morphology of the galaxy 
distribution. In other words, the cosmic geometry apparently 
implies a `geometrical biasing'' effect, qualitatively different 
from the more conventional ``peak biasing'' picture\cite{K84}.

\vfill
\end{document}